\documentclass[a4paper,fleqn,usenatbib]{mnras}

\usepackage{newtxtext,newtxmath}

\usepackage[T1]{fontenc}
\usepackage{ae,aecompl}

\usepackage{graphicx}	
\usepackage{amsmath}	
\usepackage{amssymb}	

\def\L{\mathcal{L}}

\title[Fe\,{\normalsize \textit{I}} in the atmosphere of WASP-121b]
{Detection of Fe {\Large \textbf{I}} in the atmosphere of the ultra-hot Jupiter WASP-121b, and a new likelihood-based approach for Doppler-resolved spectroscopy}

\author[N. P. Gibson et al.]{
Neale P. Gibson$^{1}$\thanks{E-mail: n.gibson@tcd.ie},
Stephanie Merritt$^{2}$,
Stevanus K. Nugroho$^{2}$,
Patricio E. Cubillos$^{3}$,\newauthor
Ernst J. W. de Mooij$^{2}$,
Thomas Mikal-Evans$^{4}$,
Luca Fossati$^{3}$,
Joshua Lothringer$^{5}$,\newauthor
Nikolay Nikolov$^{5}$,
David K. Sing$^{5}$,
Jessica J. Spake$^{6,5}$,
Chris A. Watson$^{2}$ and \newauthor
Jamie Wilson$^{2}$
\smallskip
\\
$^{1}$School of Physics, Trinity College Dublin, Dublin 2, Ireland\\
$^{2}$Astrophysics Research Centre, School of Mathematics and Physics, Queens University Belfast, Belfast BT7 1NN, UK\\
$^{3}$Space Research Institute, Austrian Academy of Sciences, Schmiedlstrasse 6, A-8042 Graz, Austria\\
$^{4}$Department of Physics, Massachusetts Institute of Technology, Cambridge, MA 02139, USA\\
$^{5}$Department of Earth and Planetary Sciences, Johns Hopkins University, Baltimore, MD, USA\\
$^{6}$Physics and Astronomy, Stocker Road, University of Exeter, Exeter, EX4 3RF, UK\\
}

\date{Accepted 2020 January 15. Received 2020 January 15; in original form 2019 November 11.}

\pubyear{2020}

\begin{document}
\label{firstpage}
\pagerange{\pageref{firstpage}--\pageref{lastpage}}
\maketitle

\begin{abstract}

High-resolution Doppler-resolved spectroscopy has opened up a new window into the atmospheres of both transiting and non-transiting exoplanets. Here, we present VLT/UVES observations of a transit of WASP-121b, an `ultra-hot' Jupiter previously found to exhibit a temperature inversion and detections of multiple species at optical wavelengths. We present initial results using the blue arm of UVES ($\approx$3700\,--\,5000\,\AA), recovering a clear signal of neutral Fe in the planet's atmosphere at >8$\,\sigma$, which could contribute to (or even fully explain) the temperature inversion in the stratosphere. However, using standard cross-correlation methods, it is difficult to extract physical parameters such as temperature and abundances. Recent pioneering efforts have sought to develop likelihood `mappings' that can be used to directly fit models to high-resolution datasets. We introduce a new framework that directly computes the likelihood of the model fit to the data, and can be used to explore the posterior distribution of parameterised model atmospheres via MCMC techniques. Our method also recovers the physical extent of the atmosphere, as well as account for time- and wavelength-dependent uncertainties. We measure a temperature of $3710^{+490}_{-510}$\,K, indicating a higher temperature in the upper atmosphere when compared to low-resolution observations. We also show that the \ion{Fe}{i} signal is physically separated from the exospheric \ion{Fe}{ii}. However, the temperature measurements are highly degenerate with aerosol properties; detection of additional species, using more sophisticated atmospheric models, or combining these methods with low-resolution spectra should help break these degeneracies.
\end{abstract}




\section{Introduction}

Transmission and emission spectroscopy are key techniques for the characterisation of exoplanet atmospheres \citep{Seager_2000,Brown_2001}. Transmission spectroscopy aims to measure the starlight that filters through the upper atmosphere of a planet as it passes in front of its parent star, and enables measurements of the composition and thermal structure of the atmosphere. Emission spectroscopy effectively enables a direct spectrum of the planet's dayside and is sensitive to similar information, although at different altitudes, at different locations on the surface, and with considerably differing sensitivity to trace atmospheric constituents given the different geometry. Both techniques are highly complementary, and combined can enable a detailed picture of an exoplanet's atmosphere.

Measurements of transmission and emission spectra are traditionally done at low-resolution, requiring extremely stable time-series to measure transit/eclipse depths to a precision of $\approx$$10^{-4}$. Therefore space-based measurements have typically dominated the field using the Hubble Space Telescope and Spitzer \citep[e.g.][]{Charbonneau_2002,Pont_2008,Huitson_2012,Berta_2012,Pont_2013,Kreidberg_2014,Nikolov_2015,Sing_2016}. More recently, ground-based observations are playing an increasingly prominent role, with the use of multi-object spectrographs to perform stable time-series observations \citep[e.g.][]{Bean_2010,Bean_2011,Crossfield_2013,Gibson_2013a,Gibson_2013b,Jordan_2013,Kirk_2016,Lendl_2016,Mallonn_2016,Nikolov_2016,Nikolov_2018,Stevenson_2014b}. However, instrumental systematics have long been the major limitation of low-resolution observations \citep[e.g.][]{Sing_2009,Deming_2013,Gibson_2011,Gibson_2019}, and despite many efforts to develop robust statistical techniques  to extract low-resolution spectra \citep[e.g.][]{Carter_2009,Gibson_2012,Waldmann_2012,Gibson_2014}, many contradictary results remain \citep[e.g.][]{Gibson_2017,Gibson_2019}. 

The development of Doppler-resolved spectroscopy \citep{Snellen_2010,Brogi_2012,Rodler2012,Birkby_2013} has dramatically increased our ability to detect exoplanet atmospheres from the ground using high-resolution spectrographs.
Doppler-resolved spectroscopy uses the large Doppler shift of the planet relative to the host star and telluric absorption to isolate the signal of the planet. At high-resolution, individual atomic and molecular lines are resolved, and the large velocity of the planet results in lines shifting significantly in subsequent exposures. This enables the (relatively stationary) stellar and telluric lines to be removed before summing in the rest frame of the planet, and the planet's signal can be further boosted using cross-correlation with model templates to effectively sum up over many spectral lines. This makes the high-resolution technique particularly powerful for robustly identifying specific atomic and molecular species that can easily be confused at low-resolution. Accessing the velocity information of the planetary signal also enables an extra dimension to isolate the planet's signal from time- and wavelength-dependent systematics, arguably making this technique less sensitive to spurious signals resulting from poorly-understood instrumental systematics. {This technique has so far confirmed the presence of molecules such as CO, H$_2$O, TiO, HCN, CH$_4$ and atomic species such as Fe, Ti, Mg and Ca \citep[e.g.][]{Snellen_2010,Brogi_2012,Birkby_2013,Nugroho_2017,Hoeijmakers2019,Hawker2018,Guilluy2019,Turner2019,Yan2019}. For a recent review see \citet{Birkby2018}.}

However, Doppler-resolved spectroscopy also has significant limitations. While it is particularly efficient at identifying atomic and molecular species, it is difficult to set detection significances in a fully principled way, or compare atmospheric models as required to extract physical information on the planet's atmosphere.
The latter is obviously critical for learning about the physics and chemistry of planetary atmospheres, and the former will be even more important once we start searching for biomarkers using next-generation telescopes \citep[e.g.][]{Snellen2013}.
These limitations arise in part due to the standard data-processing methods removing the planet's continuum that limits the information content, but also because the statistical techniques required to analyse the data are relatively poorly understood.
Developing a robust statistical framework is therefore a priority for fully exploiting high-resolution observations.
Recent pioneering work by \citet{Brogi2019} has begun to place the high-resolution technique on a principled statistical framework \citep[building on earlier work by][]{Brogi_2017}, demonstrating that retrievals of temperatures and abundances are possible directly from high-resolution observations. See \citet[][]{Fisher2019} for an alternative approach based on a machine learning framework.

Here, we present high-resolution transit observations of the ultra-hot Jupiter WASP-121b \citep{Delrez2016} with the UV-Visual Echelle Spectrograph (UVES) at the Very Large Telescope \citep[VLT;][]{UVES}, which has been previously used for measuring exoplanet spectra \citep[e.g.][]{Snellen2004, Czesla2015, Khalafinejad_2017, Gibson_2019}. WASP-121b orbits a bright, V$\sim$10.5 F6V host with a period of only 1.27\,days, giving it an equilibrium temperature of over 2400\,K. Coupled with its inflated radius, it is an excellent target for transmission spectroscopy. Indeed, WASP-121b has already been observed extensively at low-resolution with HST, with a detection of water and tentative evidence for TiO \citep[][see also \citealt{Tsiaras2018}]{Evans2016}, direct measurement of a temperature inversion via water emission bands \citep{Evans2017}, evidence for VO and an unknown blue absorber (suggested to be SH) at $\lesssim$4000\,\AA\ \citep{Evans2018}, and further emission measurements that do not strongly support the detection of VO \citep{Evans2019}, but rather point to H$^-$ emission. UV observations have shown that there is an extended (and escaping) atmosphere, with detection of \ion{Fe}{ii} and \ion{Mg}{ii} during transit extending much higher than previously detected optical and near-infrared features \citep{Sing2019}, similarly to other ultra-hot Jupiters \cite[e.g.][]{Fossati2010,Fossati2013,Haswell2012}. A tentative excess absorption in NUV observations was previously reported by \citet{Salz2019} already hinting at metal absorption from an extended exosphere. In addition, eclipse observations in the $z^\prime$ band determine upper limits on the planet's albedo \citep{Mallonn_2019}. Indeed, WASP-121b is fast becoming one of the most-observed transiting planets. The optical transmission spectra in particular show that there is excess absorption bands in the blue optical, although the exact nature of these species is unknown. This makes WASP-121b an exceptional target for high-resolution Doppler-resolved spectroscopy.

In this paper we present preliminary results using the blue arm of UVES, that is able to target atomic and molecular features that may be the cause of the temperature inversion. In addition, we outline a new technique that is able to directly evaluate the likelihood function of the model template fitted to the data, which demonstrates several advantages over previous methods. In Sect.~\ref{sect:observations} we discuss the observations and basic reductions. Sect.~\ref{sect:analysis} presents the processing of the data, atmospheric models, and search for atomic and molecular species using cross-correlation techniques, as well as the description and application of a new likelihood-based approach to high-resolution observations. Finally, Sects.~\ref{sect:discussion} and \ref{sect:conclusion} discuss the results and implications.

\section{UVES Observations}
\label{sect:observations}

Two transits of WASP-121b were observed using the 8.2-m `Kueyen'  telescope (Unit Telescope 2) of the VLT with UVES: a high-resolution echelle spectrograph \citep{UVES}. Observations were taken on the nights of 2016 December 25 and 2017 Jan 3, as part of program 098.C-0547 (PI: Gibson). Both transits were observed using a `free template' with dichroic \#2 and cross-dispersers \#2 and \#4 (for the blue and red arms, respectively).  In this paper we only consider the blue arm, and leave analysis of the red arm and a comprehensive search for additional features to a future publication (Merritt et al., {\it in prep}). We used a central wavelength of 437\,nm for the blue arm. Observations were taken with image slicer \#3, to maximise both the throughput and spectral resolution, with decker height (slit length) of 10 arcseconds, giving R$\sim$80,000 from $\approx$3750 to 4990\,\AA\ over 31 spectral orders.

Both transits used an exposure time of 100 seconds, and coupled with a readout time of $\approx$24 seconds resulted in a cadence of 124 seconds. The first transit observations lasted $\approx$4.7\,hours, resulting in 137 exposures, with approximately the first 28 before ingress, 83 in-transit, and 26 after egress. Guiding was lost for exposures 70--72, and these were clipped from the subsequent data analysis. 
The second transit observations lasted $\approx$5.0\,hours, resulting in 143 exposures, with 39, 80 and 24 before ingress, during transit, and after egress, respectively.

Data were analysed using a custom pipeline written in {\sc Python} as outlined in \citet{Gibson_2019}, which performed basic calibrations (bias/overscan subtraction), and extracted time-series spectra for each spectral order. Here, we made further modifications to perform aperture and optimal extraction in image slicer mode, which required combining the flux from five spectral traces for each order.
We used the ESO/UVES pipeline ({version 5.7.0}) to produce bias and flat-field frames, as well as perform the initial wavelength calibration and determine the trace positions for the spectral orders. Our pipeline enables much finer control over the final data products, e.g. enabling extraction of the raw flux from each order before re-sampling. We tested various extraction methods, including aperture extraction and optimal extraction, and in the end used a simple aperture extraction with aperture width of 60 pixels which comfortably contains all five traces for each order. However, we first performed cosmic ray rejection using an optimal extraction algorithm to construct the virtual PSF and replace outlying pixels. Due to the five spectral traces filling the length of the slit, we did not perform background subtraction, which is negligible for our observations.

We proceeded to use the raw flux from the extracted orders, rather than merge orders together via resampling combined with flat-fielding and blaze correction, as the response changes as a function of time and cannot be corrected via flat-fielding. We also checked the stability of the wavelength solution by cross-correlation of the spectral time-series using a stellar template \citep{Husser2013} after {applying a high-pass filter to continuum-subtract} the orders, finding the total drift was $\lesssim$\,1.3 and 1.6 km/s over each night, respectively, significantly smaller than a resolution element ($\approx$3.8\,km/s with R$\sim$80,000) but larger than the barycentric corrections.
However, we found that the order-to-order range in the wavelength solution was $\approx$3\,km/s, likely a result of inaccuracies in the wavelength solution determined from the pipeline. We later correct for the order-to-order drift, but do not attempt to correct order-to-order dispersion, which we keep in mind for our interpretation of the measured planet velocities.
The throughput of the instrument peaks at the bluer end of the wavelength range, with typical S/N per spectral element of $\approx$30--40 (at the centre of the orders) for the 1st night. Due to variable throughput (due to both weather and pointing issues), the S/N typically varies from $\approx$5--40 for the 2nd night with lower count rates during the transit. Given the significantly lower data quality of the 2nd night, we only present analysis for the 1st transit, as we did not have the S/N or stability to confirm our results. {Finally, we computed the Barycentric Julian Date for each observation (using {\sc Astropy.Time} routines).}

{We first removed outliers in each order by subtracting a simple model for the data, constructed from the outer product of the median spectrum (i.e. median over time) and median light curve (i.e. median over wavelength) divided by the data mean (to correct the scaling).} We then fitted a 10th order polynomial to each residual spectrum, and replaced any 5\,$\sigma$ outliers (initially assuming Poisson noise from our extraction pipeline) by the value of the polynomial fit, before adding back the {simple model for each order}.

Given the variations in flux both as a function of wavelength and time, it is important to have realistic estimates for the uncertainties to optimise the extraction of information from the spectra. Poisson-noise estimates from count rates and read-noise can bias fits to the residuals after removing the stellar spectra (fundamental to the high-resolution technique), by imposing that negative values of the residuals (i.e. lower count rates) will always have lower uncertainties than positive values. It is therefore important to remove this `noise' from the uncertainty determination.

We constructed initial guesses for our uncertainties by assuming Poisson noise and a single `background' value (which also accounts for read-noise) of the form $\sigma_{i}=\sqrt{aF_i+b}$, where $F_i$ is the flux for a given time and wavelength. 
We did not attempt to account for additional contributions within telluric features or from wavelength-dependent background values, which is likely to be minimal at these wavelengths.
We then subtracted a 5th-order Principle Component Analysis (PCA) decomposition for each order (which removed the stellar spectrum) to find residuals $R_i$, and determined values for $a$ and $b$ (and therefore uncertainties $\sigma_i$) by optimising a Gaussian log likelihood function (after removing constant terms):
\[
\ln\L(a,b) = -0.5\sum_i \left(\frac{R_i}{\sigma_i}\right)^2 - \sum_i\ln\sigma_i.
\]
This was done separately for each order. Finally, we reconstructed the uncertainties from the estimated $\sigma_i$ using a 5th-order PCA, which removed the `noise' from the estimate of the uncertainties, and therefore any bias in fitting the residuals. We found that without this final step, the amplitude of the best-fit model fitted to the data was biased towards smaller values (checked via injection tests). {We note that this procedure was to obtain a reasonable estimate of the uncertainties, and our results are not particularly sensitive to the exact number of PCA components used.}


\section{Analysis}
\label{Analysis}
\label{sect:analysis}

\subsection{Pre-processing the spectra: removal of stellar and telluric features}
\label{sect:preprocessing}

\begin{figure*}
\centering
\includegraphics[width=180mm]{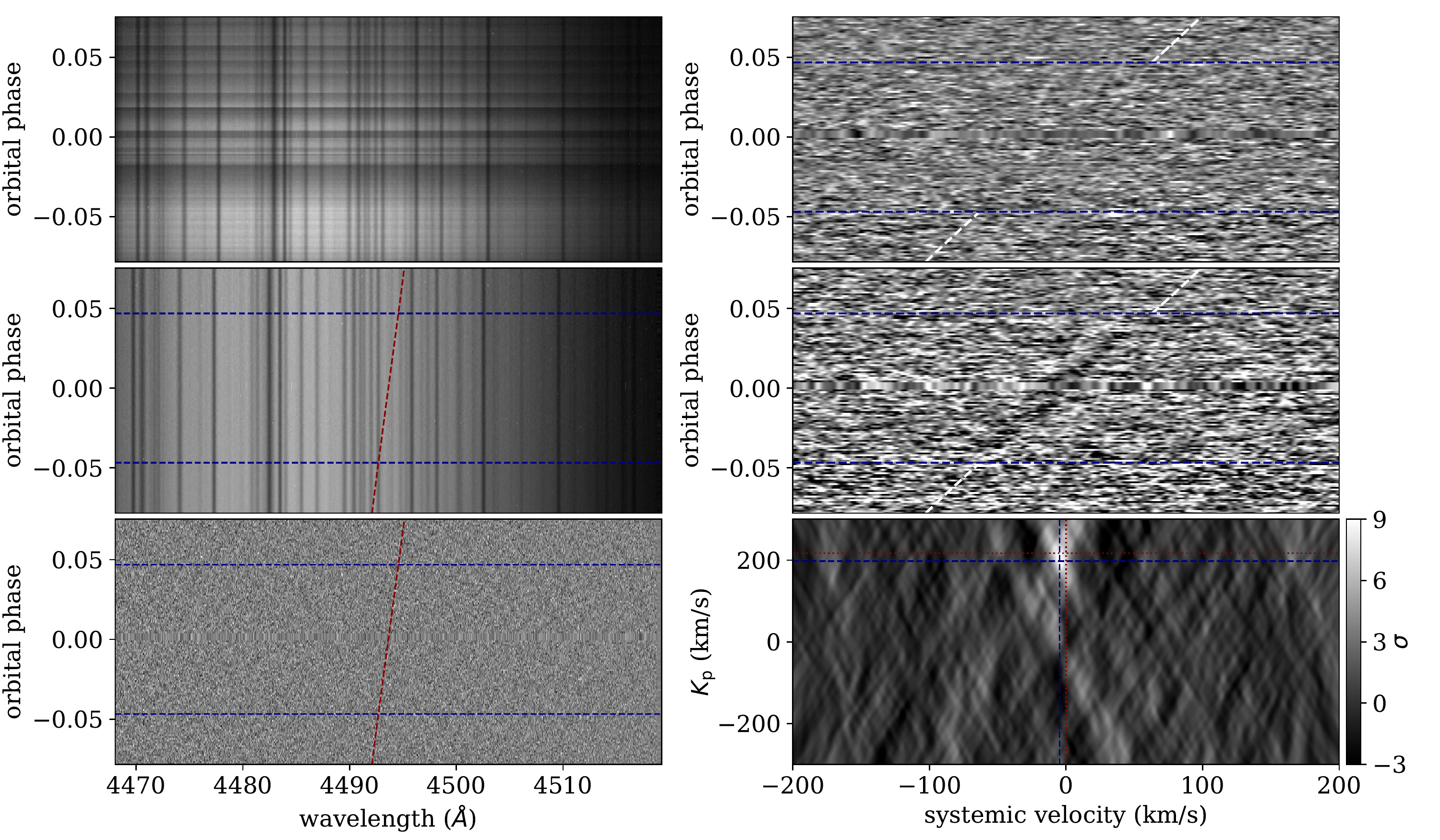}
\caption{Demonstration of the data-processing of a single order and detection of \ion{Fe}{I} in WASP-121b. Left: Data-processing of a single raw order (top), after wavelength-shift and blaze-correction (middle), and after division through by the {\sc SysRem} model (shown weighted by the uncertainties, bottom). The horizontal dashed lines show the positions of ingress and egress, and the diagonal line shows the approximate velocity shift of WASP-121b. 
Right: Cross-correlation of the data, including a single order after cross-correlation with a model template (top), summed cross-correlation of the bluest 13 orders (middle), and velocity-summed cross correlation map (bottom) as a function of planet orbital velocity ($K_{\mathrm p}$). For the upper two plots the colormap was inverted for clarity (i.e. dark shows positive correlation).
The dashed line marks the peak of the detection, and the dotted lines mark the expected solution. The colour bar shows the detection significance weighted on the standard deviation of the cross-correlation map outside the peak, showing a clear detection of \ion{Fe}{I} in the atmosphere of WASP-121b (see text).
}
\label{fig:processing}
\end{figure*}

We begin by using methodology common to high-resolution searches for molecular features via cross-correlation \citep[e.g.][]{Snellen_2010,Brogi_2012,Birkby_2013}, in order to disentangle the Doppler-shifted planetary transmission spectrum from the (effectively) static stellar and telluric features. An example of the raw spectra and subsequent pre-processing steps on a single order are shown in Fig.~\ref{fig:processing}.

The first step is to remove blaze variations, i.e. time-dependent changes in the shape of the spectral orders due to varying throughput of the instrument.
We followed the procedure outlined in \citet{Gibson_2019}. This consisted of dividing each spectrum through by a master spectrum (the median of the time-series for each order), smoothing each result with a median and then Gaussian filter (with a width/standard deviation of 11 and 50 pixels, respectively), and dividing each original spectrum through by the resulting smoothed blaze distortion. This does not remove the blaze function, but rather places the spectra on a `common' blaze. Due to the instability of the blaze correction at the blue end of each order, we removed the first 600 pixels from the analysis. We also clipped the final 60 pixels from each order. {This left a total of 2340 pixels in each spectrum out of a total of 3000 pixels in the raw spectra.}

We then apply a wavelength correction to the spectral time-series. This is derived from the cross-correlation of the data with a stellar template as described in Sect.~\ref{sect:observations}, and we interpolate the spectra to the stellar rest frame after smoothing the measured velocity shifts with a Gaussian process (GP) with standard squared exponential kernel \citep[see][for an introduction to GPs in the context of exoplanet time-series analysis]{Gibson_2012}. This process did not consider potential errors in the dispersion for each order.
Note that this corrects for the barycentric velocity variations and the systemic velocity, but will also fit the Rossiter-McLaughlin (RM) effect. The RM effect for WASP-121b has been measured to have an amplitude of $\sim$100\,m/s (with $v\sin i = 13.6$\,km/s), and is on a nearly-polar orbit with projected obliquity of 258\,deg \citep{Delrez2016}. The velocity amplitude of the RM effect is substantially smaller than the change in velocity of WASP-121b during transit ($\gtrsim$100\,km/s from ingress to egress), so it will have minimal impact on our final results. In principle it could impact the removal of the stellar features due to small misalignments in the stellar spectra; nonetheless, accounting for the RM effect in the velocity correction would also leave small variations in the line shapes during transit, and a full treatment would require accounting for the variations in the line shapes. Either way, we note that this cannot mimic a planetary signal at the velocity of WASP-121b.

\begin{figure}
\centering
\includegraphics[width=86mm]{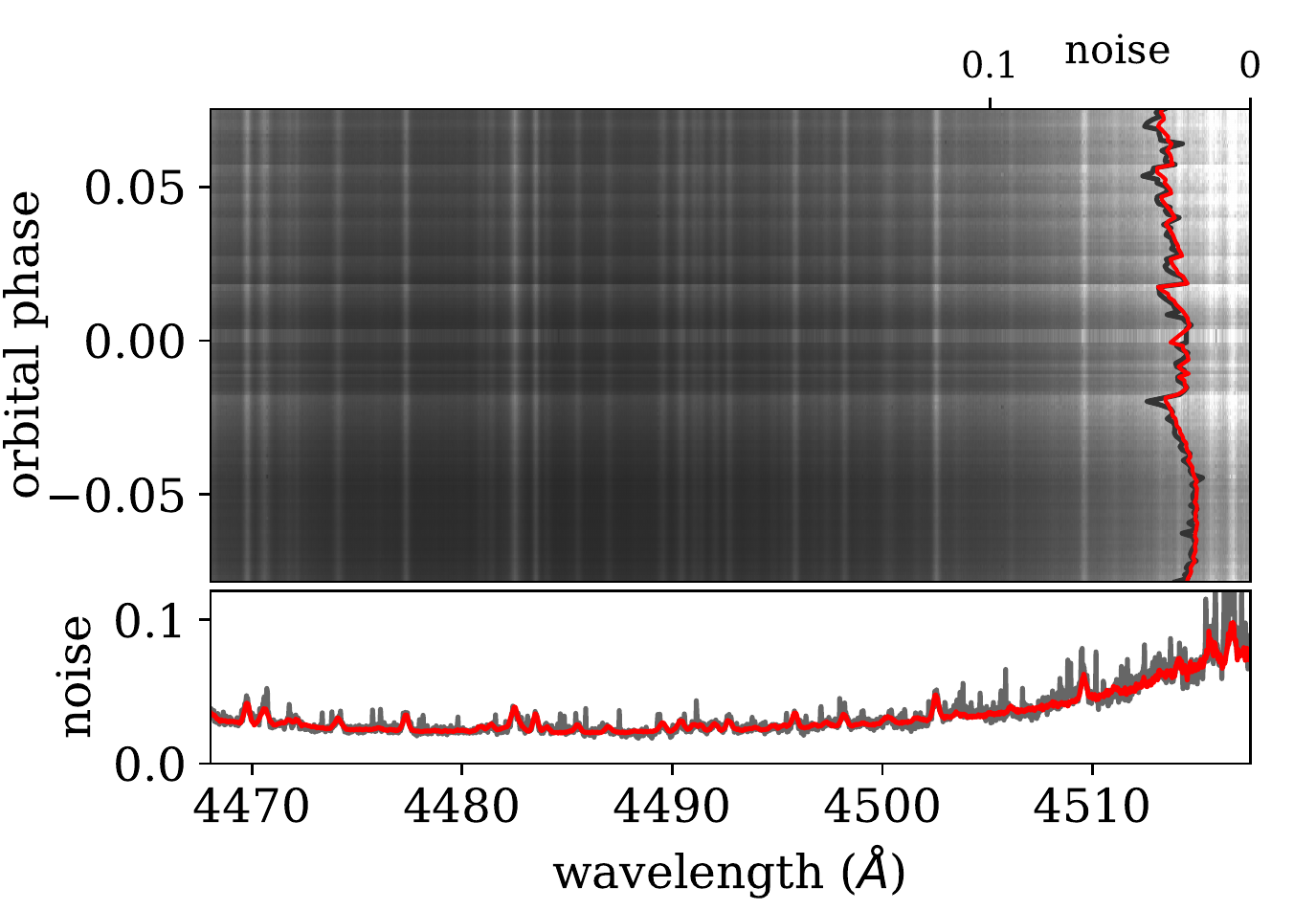}
\caption{Visualisation of the noise in a single processed order. The upper panel shows the calculated uncertainty after pre-processing (i.e. the uncertainty of the residual array shown in the lower left panel of Fig.~\ref{fig:processing}), demonstrating the strong time- and wavelength-dependence of the noise. The lower panel shows the standard deviation over time for each spectral pixel (grey), and the mean of the calculated uncertainty. The equivalent wavelength-averaged values are shown in the upper panel.}
\label{fig:noise}
\end{figure}

The next step is to remove the stellar and telluric features from the time-series. Similarly to \citet{Gibson_2019}, we applied the {\sc SysRem} algorithm \citep{sysrem} which was first demonstrated on high-resolution spectroscopy by \citet{Birkby_2013}, and has since become a standard technique \citep[e.g.][]{Birkby_2017,Esteves_2017,Nugroho_2017,Deibert2019}. First, we apply {\sc SysRem} to each order in the standard way, and determine a model representation of the 2D data array (time vs wavelength) as an outer product of two column vectors, before subtracting from the data and determining the next {\sc SysRem} iteration. However, rather than using the end product (residuals from repeated subtractions), we sum the {\sc SysRem} models for each iteration to determine a global model for the order, and {divide} the dataset through by the model before subtracting one. This procedure preserves the relative depths of the planet's transmission spectrum, while allowing us to model the raw measured flux directly. The uncertainties are also divided through by the same stellar and telluric model. Fig.~\ref{fig:processing} shows the final processed spectra and processing steps used for a single order, and Fig.~\ref{fig:noise} shows the corresponding uncertainties.

\subsection{Model transmission spectra}
\label{sect:modelling}

In order to search for atomic and molecular features using the high-resolution cross-correlation technique, we first require models of the planet's atmosphere to create cross-correlation templates. We first produce absorption cross-sections for a range of species (including \ion{Fe}{I} and \ion{Fe}{II}) and temperatures using the {\sc Helios-K} code along with the Kurucz atomic line lists \citep{Grimm2015,Kurucz2018} assuming local thermal equilibrium. This is a reasonable assumption deep in the atmosphere, but may not be reliable for considering species in the exosphere. We only consider thermal broadening using a line-wing cutoff of $10^8$ times the Lorentz width (i.e. effectively without truncation), and create cross-sections with a resolution of $R>2,000,000$ from temperatures of 1500 to 4500\,K in 50\,K steps.

Armed with our cross-sections, we compute model templates for WASP-121b following Equation 12 of \citet{HengKitzmann2017} which provides an approximate transmission spectrum for an isothermal atmosphere (and with isobaric cross-sections). We modify the equation to depend on cross-section per atom/molecule $\sigma(\lambda)$, and to incorporate multiple species indexed by $j$, computing the effective radius as:
\[
R(\lambda)=R_{0}+H\left[\gamma+\ln \left(\frac{P_{0}}{mg} \sqrt{\frac{2 \pi R_{0}}{H}}\right)\right] + H \ln\sum_j \chi_j \sigma_j(\lambda).
\]
Here, $H$, $g$ and $m$ are the scale height, surface gravity and mean molecular mass of the atmosphere, $R_{0}$ and $P_{0}$ are the reference radius and pressure, $\gamma = 0.56$ is a dimensionless constant, and $\chi_j$ are the volume mixing ratios. Only the final term varies in our model, which enables fast determination of the model spectrum. For a given temperature ($T_\mathrm{atmos}$), we linearly interpolate the cross-section from our pre-computed grid.

We also account for scattering in our simple model by including both Rayleigh scattering and a cloud deck. For Rayleigh scattering, we use the cross-section of \citet{Lecavelier_2008b}, and allow for an arbitrary increase in abundance ($\chi_\mathrm{ray},~$relative to H$_2$) to increase the scattering strength (e.g. due to larger scattering particles). For the cloud deck, we simply truncate the model at a given pressure level ($P_\mathrm{cloud}$).
Another important potential source of scattering in ultra-hot Jupiters is H$^-$ \citep[see][for a detailed discussion in the context of WASP-121b]{Parmentier2018}, however we do not consider it here in order to avoid further degeneracy with the cloud deck and Rayleigh scattering. Our final model (for a single species) therefore contains the abundance, temperature, cloud-deck pressure, and Rayleigh scattering strength (relative to H$_2$). For each model we convert to differential flux ($\Delta F$), by computing the (negative of the) square of the planet-to-star flux ratios. We also separately compute the scattering model (clouds plus Rayleigh scattering only), and subtract this from our final model. Finally, we perform a subtraction of any continuum that arises from dense, overlapping lines. {Unless otherwise stated, we subtracted a (Gaussian-smoothed) maximum filter (both with a window of 50 resolution elements or $\approx$$0.07-0.12$\,\AA) from the model.} The continuum is removed from the model as the pre-processing steps outlined in Sect.~\ref{sect:preprocessing} will remove any slowly varying features from the data. We found that our results were not particularly sensitive to the choice of filter widths.

We use the system parameters from \citet{Delrez2016}; however, given that we are subtracting the continuum we are not particularly sensitive to the system parameters, and it is irrelevant for the initial cross-correlation analysis. To calculate the scale height, we fix the temperature to 2800\,K \citep[approximately the measured temperature at the top of the atmosphere from][]{Evans2019} and assume a Jupiter-like composition. This results in a scale height of $H$\,=\,960\,km. {In principle, we could re-calculate the scale height using the temperature assumed for the \ion{Fe}{I} cross-section ($T_\mathrm{atmos}$). However, we prefer to fit directly for a scale factor of the model (see Sect.~\ref{sect:mapping}) and interpret it afterwards, in particular due to the detection of the extended \ion{Fe}{II} signal which demonstrates that a hydrostatic atmosphere is not valid for modelling species in the upper atmosphere \citep{Sing2019}.}

Given the degeneracy between reference pressure, radius, and abundance, \citep[e.g.][]{Benneke2012, HengKitzmann2017}, i.e. one can vary the abundance of a species as well as the reference radius and produce an identical transmission spectrum (assuming a fixed scale height), we fix the abundance of the absorbing species, and vary the strength of the Rayleigh scattering and pressure of the cloud deck. This degeneracy in principle can be broken by measuring a pressure-sensitive feature \citep[e.g. pressure broadening or H$_2$ Rayleigh scattering;][]{Benneke2012}, but we ignore pressure broadening here, and our data-processing removes any sensitivity to the continuum (other than varying the line-strengths visible above the continuum, but this is also dependent on temperature).

\begin{figure}
\centering
\includegraphics[width=80mm]{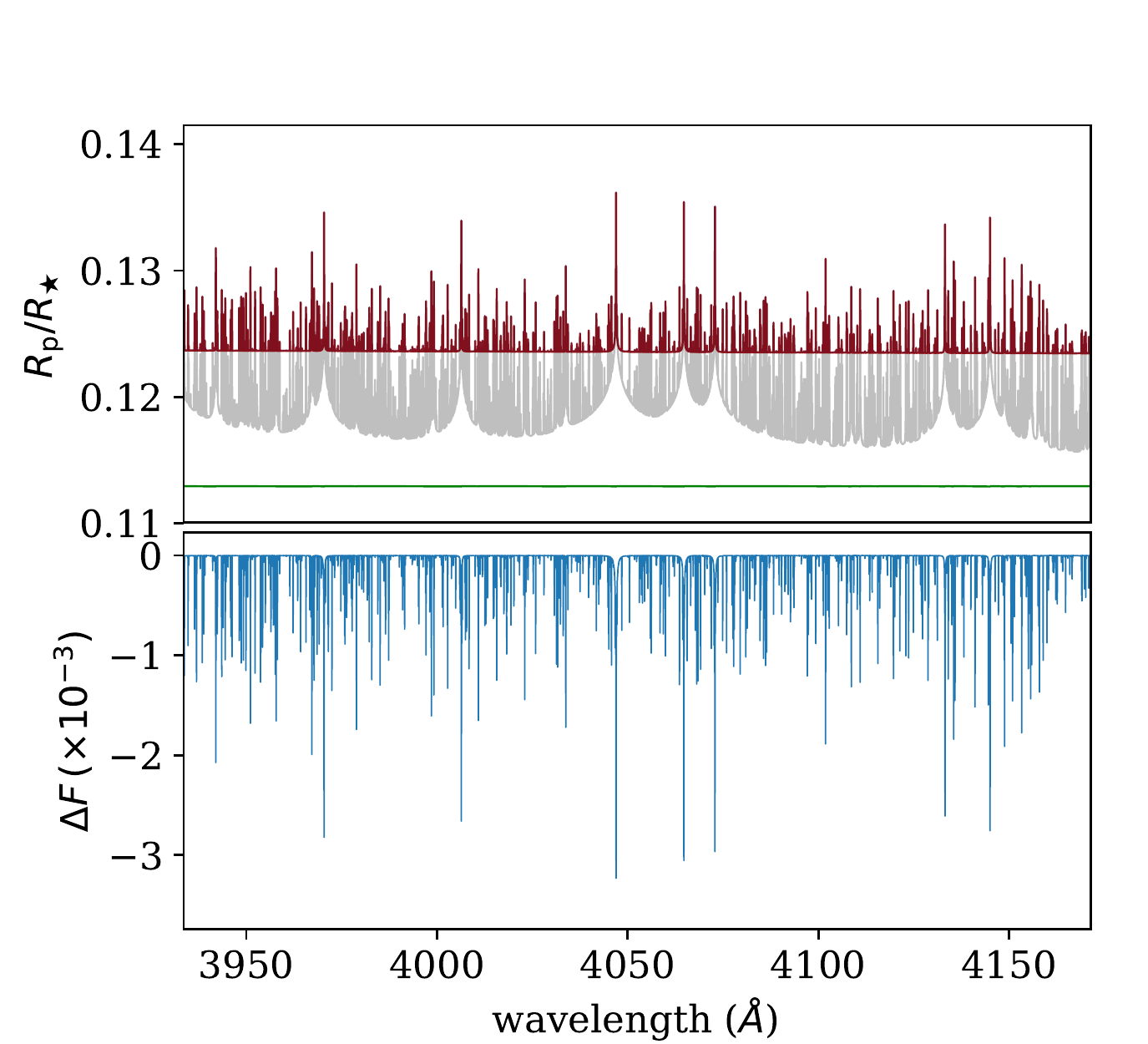}
\caption{Examples of the model transmission spectra for WASP-121b using \ion{Fe}{I}, Rayleigh scattering and a cloud deck. Top: The grey model shows the scattering-free model (effectively scaled log opacity of \ion{Fe}{I}). The full models (i.e. with scattering) computed using both the fast approximation (dark red) and a 50-layer radiative-transfer model (dark blue) are over-plotted. These are indistinguishable, and the green line shows the difference (fast-approximation minus layered model, with arbitrary offset) between the two on the same scale. Bottom: Final model template after continuum subtraction in units of flux.}
\label{fig:models}
\end{figure}

Initially, we generate simple models for the purposes of detecting species using an (arbitrary) abundance of 10$^{-6}$, and explore a range of scattering parameters that truncate the model at various pressures. Later, we fit our parameterised model directly to the data (see Sect.~\ref{sect:MCMCfits}). We verify our models by comparing to a radiative transfer model after integrating over 50 isothermal layers equally spaced in log pressure, using the same cross-section at each layer, and incorporating Rayleigh scattering and an opaque cloud deck. Fig.~\ref{fig:models} shows examples of models generated using both the fast analytic model, and a `layered' radiative transfer model, clearly demonstrating that the fast model is sufficient for an isothermal, well-mixed atmosphere that ignores pressure-broadening.

Our final step is to Doppler shift the models (via linear interpolation) according to the planet's velocity $v_\mathrm{p}$, given by:
$$
v_\mathrm{p} = v_\mathrm{sys} + K_\mathrm{p} \sin(2\pi\phi),
$$
where $v_\mathrm{sys}$ is the systemic velocity, $K_\mathrm{p}$ is the velocity amplitude of the planet's orbit, and $\phi$ is the orbital phase of the planet (with $\phi$=0 during mid-transit). Note that the spectra have already been shifted to the stellar rest frame, meaning that the barycentric velocity shift has been removed, and the expected $v_\mathrm{sys}$ is zero. The velocity shift is applied in two ways; either by Doppler-shifting the model to a pre-defined range of velocity offsets, computing the cross-correlation as a function of $v_\mathrm{sys}$ for each time, before shifting the cross-correlation function to the rest velocity of the planet and summing; or alternatively Doppler-shifting the model for each planet velocity in the time-series for a given set of  $v_\mathrm{sys}$ and $K_\mathrm{p}$. The method we use depends on whether we are producing a full cross-correlation (or likelihood map), or directly fitting the models to the data, which is discussed in more detail in the following sections.

\subsection{A new likelihood `mapping' for high-resolution observations}
\label{sect:mapping}

In addition to applying the standard cross-correlation methods to our data, we also directly compute a full likelihood function
inspired by the approach of \citet{Brogi2019}. This enables us to explore and compare different model templates using principled statistical techniques. Here, we show that we can compute a likelihood map directly from the cross-correlation map, as well as quickly compute the log likelihood for use in model fitting. We also provide a generalised version of the likelihood mapping derived in \citet{Brogi2019}, which has the added advantage of {explicitly} accounting for time- and wavelength-dependent uncertainties.

We start with a standard Gaussian likelihood function, with uncertainties that vary in time and wavelength:
$$
\L = \prod_{i=1}^N \frac{1}{\sqrt{2\pi\sigma_i^2}} \exp\left(-\frac{1}{2}\frac{ (f_i -m_i(\bmath\theta))^2}{\sigma_i^2} \right),
$$
where $r_i = f_i -m_i(\bmath\theta)$ are the residuals from the data $f_i$ and model $m_i$, $\sigma_i$ are the uncertainties, $i$ indexes over both time and wavelength (and spectral orders), and $N$ is the number of data points. $\bmath\theta$ is the vector of parameters of the model, including the atmospheric model parameters and orbital and systemic velocities.
The fact that we allow for time- and wavelength-dependent uncertainties is the most important distinction between our approach and that of \citet{Brogi2019} at this point. {However, we note that considering time- and wavelength-dependent uncertainties is not novel for high-resolution analysis, and this can be accounted for in various ways including during the pre-processing of the data \citep[see e.g.][]{Snellen_2010,Brogi_2012}.}

Next, we introduce a scaling term for the white noise $\beta$, which enables us to preserve the time- and wavelength dependence of the noise, but also account for the fact that we may not have the right scaling. This has long been standard procedure in transit modelling \citep[e.g.][]{Winn_2008,Gibson_2008}, where we often fit for the noise in the data rather than impose it from photon statistics \citep[and is also of fundamental importance in systematics modelling, e.g.][]{Carter_2009,Gibson_2012}. An important distinction for high-resolution observations is that noise cannot be approximated as homoskedastic in nature, and large variations are present in both wavelength (due to strong stellar and telluric lines, as well as the instrument response, e.g. see Fig.~\ref{fig:noise}) and normally in time (due to slit/fibre losses and weather). We could in principle also enable a more complicated scaling function (e.g. time/wavelength dependent scaling), but we stick with a single scaling term which enables us to account for heteroskedastic noise in the simplest way possible.

We also introduce a scale factor for the model $\alpha$, to account for any uncertainty in the scale of the model. \citet{Brogi2019} assume this to equal one, and state that this is the maximum likelihood solution. We find that the maximum likelihood solution for $\alpha$ depends on the scale of both the model and data. Nonetheless, in practice setting $\alpha$\,=\,1 assumes one knows the scale height of the atmosphere, or that it is incorporated into the parameterisation of the model. {This is of course the standard approach taken for general model fitting, and the one adopted by \citet{Brogi2019}. We note that recent work by \citet{Gandhi2019} has explored fitting for an $\alpha$-like term using the \citet{Brogi2019} framework.
We choose to fit for $\alpha$ as a free parameter, in part because we are searching for features that are potentially in the exosphere of the planet, where we cannot assume a hydrostatic atmosphere.
{As we will discuss later, another advantage of fitting for $\alpha$ directly is that we can easily recompute the likelihood as a function of $\alpha$ without recomputing the cross-correlation map, effectively getting an extra parameter `for free'.}
We discuss the physical interpretation of $\alpha$ later.}
Our likelihood function therefore becomes:
$$
\L = \prod_{i=1}^N \frac{1}{\sqrt{2\pi(\beta\sigma_i)^2}} \exp\left(-\frac{1}{2}\frac{(f_i-\alpha m_i)^2}{(\beta\sigma_i)^2} \right),
$$
where we now drop explicit reference to $\bmath\theta$. In practice we compute the log likelihood:
$$
\ln\L = -\frac{N}{2}\ln2\pi-N\ln\beta-\sum_{i=1}^N\ln\sigma_i-\frac{1}{2} \chi^2,
$$
where we define $\chi^{2}$ as:
$$
\chi^2 = \sum \frac{(f_i-\alpha m_i)^2}{(\beta\sigma_i)^2},
$$
and the summation is hereafter implied over $i$. Typically any constant terms are dropped from the computation.
We can finally expand $\chi^2$ similarly to \citet{Lockwood2014} and \citet{Brogi2019}, but preserving dependence on $\alpha$ and $\beta$:
$$
\chi^2 = \frac{1}{\beta^2}\left[\sum  \frac{f_i^2}{\sigma_i^2} + \alpha^2 \sum\frac{m_i^2}{\sigma_i^2} - 2\alpha \sum \frac{f_i m_i}{\sigma_i^2} \right],
$$
and can identify the optimally-weighted cross-correlation function (CCF) as the final term:
$$
\mathrm{CCF} = \sum\frac{f_i m_i}{\sigma_i^2}.
$$
These equations enable us to compute the log likelihood or $\chi^2$ efficiently from the CCF function. The first term is a constant for the dataset, and the second term must be computed for each model (and for every Doppler shift). Once these have been evaluated one can easily compute the likelihood map as a function of $\alpha$ {without recomputing the CCF, highlighting an advantage of fitting for $\alpha$ directly.}
Alternatively we can optimise the model parameters using the log likelihood or $\chi^2$ as a merit function.
Our options for $\beta$ are either to scale it so that the reduced $\chi^2$ is one (according to the best fit model), or alternatively allow $\beta$ to be a free parameter and marginalise over it.
If $\beta$ is treated as a free parameter, then the full log likelihood function must be used, as $\beta\rightarrow\infty$ if minimising $\chi^2$. We note that $\beta$ is not required for finding the maximum likelihood solution, but is important for defining the shape of the merit function -- i.e. for determining uncertainties in model parameters.

We can instead follow a similar procedure to \citet{Brogi2019}, and introduce an alternative likelihood mapping. They choose to `null' a global value of the uncertainties $\sigma$ by setting the partial derivative of the likelihood with respect to $\sigma$ to zero. Note that they did not use a scale factor, $\beta$, but rather used a global value for $\sigma$ to remove the dependence on the value of the uncertainties, which does not account for time- and wavelength-dependent noise. In our case, we preserve the relative uncertainties, but follow the same procedure for the scale factor, i.e. `nulling' the partial derivative with respect to $\beta$:
\[
\begin{split}
\frac{\partial\ln\L}{\partial\beta} &= -\frac{N}{\hat\beta} + \frac{1}{\hat\beta^3}\sum\frac{r_i^2}{\sigma_i^2} = 0\\
\text{and rearranging to get:}\\
\hat\beta^2 &= \frac{1}{N} \sum\frac{r_i^2}{\sigma_i^2},
\end{split}
\]
where $\hat\beta$ represents the maximum likelihood estimate for $\beta$. We then substitute this into the log likelihood equation to get a more general version of Equation~9 in \citet{Brogi2019}:
\[
\begin{aligned}
\ln\L &= -N\ln\sqrt{ \frac{1}{N}\sum\frac{r_i^2}{\sigma_i^2} }\\
&= -\frac{N}{2}\ln \left[ { \frac{1}{N} \left ( \sum\frac{f_i^2}{\sigma_i^2} + \alpha^2 \sum\frac{m_i^2}{\sigma_i^2} - 2\alpha \sum \frac{f_i m_i}{\sigma_i^2} \right ) }  \right]
\end{aligned}
\]
or more elegantly:
$$
\ln\L = -\frac{N}{2}\ln \frac{\chi^{2}}{N},
$$
where $\chi^{2}$ no longer depends on $\beta$ and is defined as:
$$
\chi^{2} = \sum_{i=1}^N \frac{(f_i-\alpha m_i)^2}{\sigma_i^2}.
$$

This version of the log likelihood automatically optimises the noise scale factor $\beta$ for each model (and value of $\alpha$), with the caveat that the value of $\beta$ (and hence the uncertainties) change with the model. Strictly speaking, this is not desirable behaviour, but is a convenient approximation to reduce the dimensionality of the likelihood space which is generally expensive to compute. Our preference is to allow $\beta$ to be a free parameter in model fits, and to set the reduced $\chi^2=1$ (for the best-fit model) when producing likelihood maps. However, we later show that the choice of either likelihood mapping is not particularly important, given the huge amount of data points that enables accurate determination of $\beta$. 

Our method is in practice the statistically correct approach to fitting models to data assuming that the scaling of the uncertainties is correct (i,e. assuming the uncertainties can be given by $\beta\sigma_i$). However, this of course also assumes that the model is the correct one to explain the data, but in practice the filtering of the data with {\sc SysRem} alters the underlying model. This is a difficult problem to solve, and depends on the method used to remove the stellar and telluric features. Our solution is to apply filtering to the data as outlined in Sect.~\ref{sect:modelling}.
In theory, we should also include all species and sources of opacity in the model atmosphere when fitting, otherwise the model will not provide a good fit to the data. Neglecting this will in principle result in a worse fit to the data, and the corresponding increase in $\beta$ will produce more conservative errors in parameter fits.

\subsection{Detection of \ion{Fe}{I} via cross-correlation}
\label{sect:detection_crosscorr}

We perform a search for various species in the atmosphere of WASP-121b that have significant absorption in the spectral range of UVES, including \ion{Fe}{I}, \ion{Fe}{II}, TiI and TiII using an arbitrary range of cloud-deck pressures. However, we only find evidence for \ion{Fe}{I}, which we describe in detail here. We leave a complete inventory of species \citep[e.g.][]{Hoeijmakers2019} in the planet to future work.

We first pre-processes our data using 15 iterations of {\sc SysRem} as discussed in Sect.~\ref{sect:preprocessing}. The relatively high-number of iterations is required to fully remove the stellar \ion{Fe}{I} signal. {This was determined by visually inspecting the data, but we note that our results are not significantly changed by decreasing or increasing the number of {\sc SysRem} iterations (we checked our signal detection using both 10 and 20 iterations).}
We proceed by generating models containing \ion{Fe}{I}, Rayleigh scattering and a cloud deck as discussed in Sect.~\ref{sect:modelling} using a range of temperatures and scattering parameters. We begin by exploring temperatures ranging from 1500 to 4500\,K in steps of 50\,K, with Rayleigh scattering from H$_2$ only, and a cloud deck set at pressures of 10$^{-1}$, $5\times10^{-2}$ and 10$^{-2}$ bar. We report results for the best fit template model, as determined in Sect.~\ref{sect:CCFdetection}, but the detection of \ion{Fe}{I} is significant for a wide range of templates.

We next Doppler shift the atmospheric model to the wavelength grid of the data (via linear interpolation) using a range of systemic velocities from -200 to 200\,km/s, in steps of 0.4\,km/s (finer than the pixel scale of the data). We then multiply each of these models by each spectrum of our time series, after weighting each data point by its variance, and sum over wavelength to generate our optimal CCF as a function of time and $v_\mathrm{sys}$. This process is performed for every spectral order. An example for a single order is shown in Fig.~\ref{fig:processing}.

The next step is to sum over the planet's velocity curve as a function of time. We compute the planet's velocity as a function of time as described in Sect.~\ref{sect:modelling} (with $v_\mathrm{sys}$\,=\,0, as we have already corrected to the stellar rest frame), interpolate each CCF of the time-series to the rest frame of the planet, and finally sum over time. We also weight each spectra in the sum according to a transit model of WASP-121b, as the planet's signal is only present during transit, and weaker during ingress and egress. We use the equations of \citet{Mandel_Agol_2002} and assume no limb darkening, and use the ephemeris determined by \citet{Sing2019}. To characterise any systematic effects in the data as well as compute the detection significance, we repeat this step over a range of $K_\mathrm{p}$ from -300 to 300\,km/s in steps of 1\,km/s, to produce a $K_\mathrm{p}$ vs $v_\mathrm{sys}$ cross-correlation `map'.

The final procedure is to sum these maps over the spectral orders, and the resulting cross-correlation map is shown in Fig.~\ref{fig:processing}. 
{This shows a clear peak near the expected values of $K_\mathrm{p}$ \citep[$\approx$217$\pm$15\,km/s using masses and stellar velocity from][]{Delrez2016} and $v_\mathrm{sys}$ (0\,km/s), although a slight offset is detected, which we discuss in more detail in Sect.~\ref{sect:discussion}.} We first compute a rough detection significance {\citep[sometimes referred to as signal-to-noise rather than detection significance, e.g.][]{Brogi2018}} by dividing the map through by its standard deviation  taken from a $K_\mathrm{p}$ of 150 to 250\,km/s and a $v_\mathrm{sys}$ from -120 to -40 and 40 to -120\,km/s, i.e avoiding the peak of the cross-correlation map and zero $K_\mathrm{p}$ (where we expect slightly different noise properties). 
This results in a detection significance of 9.2\,$\sigma$, with no other significant peaks present in the CCF map, although the exact detection significance will slightly vary with the (arbitrarily chosen) noise regions.
We note that if we do not weight by the individual uncertainties (equivalent to assuming identical uncertainties for all times and wavelengths), the detection significance drops to 5.0\,$\sigma$. This highlights the importance of fully taking into account the heteroskedastic nature of the noise. Fig.~\ref{fig:processing} also shows the sum of the CCF functions over a subset of orders as a function of time, i.e. prior to summation over the planet's velocity curve. It is possible to identify the trace of the planet's CCF moving in time close to the expected $K_\mathrm{p}$, providing further confidence in the detection of \ion{Fe}{I} in the planet's atmosphere.

\subsection{Application of the new likelihood via CCF map}
\label{sect:CCFdetection}

While computing the CCF, it is straightforward to compute the likelihood function directly using the equations presented in Sect.~\ref{sect:mapping}. If we assume $\beta$ is fixed, the log Likelihood can simply be written as:
$$
\ln\L = -\frac{1}{2} \chi^2,
$$
after removing constant terms, and where
$$
\chi^2 = \frac{1}{\beta^2} \left[ \sum  \frac{f_i^2}{\sigma_i^2} + \alpha^2 \sum\frac{m_i^2}{\sigma_i^2} - 2\alpha \sum \frac{f_i m_i}{\sigma_i^2} \right].
$$

The final summation in the $\chi^2$ is the CCF function, and the other terms are straightforward to compute. The first term is a constant, but is required to set the value of $\beta$ from the reduced $\chi^2$, and is pre-computed. The middle term is easily computed for each model after Doppler shifting; note that this must be computed for every value of $v_\mathrm{sys}$ and each spectrum.

We proceed to compute the likelihood function directly from the CCF (initially assuming $\beta=1$), using a range of atmospheric models. For each CCF evaluation, we compute the other terms in the $\chi^2$, before summing up over the planets velocity as before. We compute the CCF for 40 values of $\alpha$ equally spaced from 0 to 2, and produce model atmospheres for a range of temperatures ranging from 1500 to 4500\,K in steps of 50\,K, and for cloud deck pressures of $10^{-1}$, $5\times10^{-2}$, and $10^{-2}$\,bar. Our final log likelihood map is therefore 5 dimensional, and depends on $K_\mathrm{p}$, $v_\mathrm{sys}$, $\alpha$, $T_\mathrm{atmos}$ and $P_\mathrm{cloud}$, with a fixed value of $\beta$.
Note that the standard CCF `map' is only a function of $K_\mathrm{p}$ and $v_\mathrm{sys}$. In principle it can be computed for many different models, but it is difficult to directly compare CCFs in a principled manner. Our likelihood maps (or equivalently posterior maps, where we have uniform priors defined over the linear or logarithmic parameter spacing) enable us to take slices (conditional distributions) or sums (marginalised distributions) over parameters in a trivial way, and explore the sensitivity of the likelihood to a range of models and therefore get best fit parameters and uncertainties. We finally determine the value of $\beta$ from the best-fitting model, so that the minimum reduced $\chi^2=1$, and re-scale the log likelihood accordingly. This is equivalent to optimising the log likelihood with respect to $\beta$ conditioned on the best fit model.

\begin{figure}
\centering
\includegraphics[width=86mm]{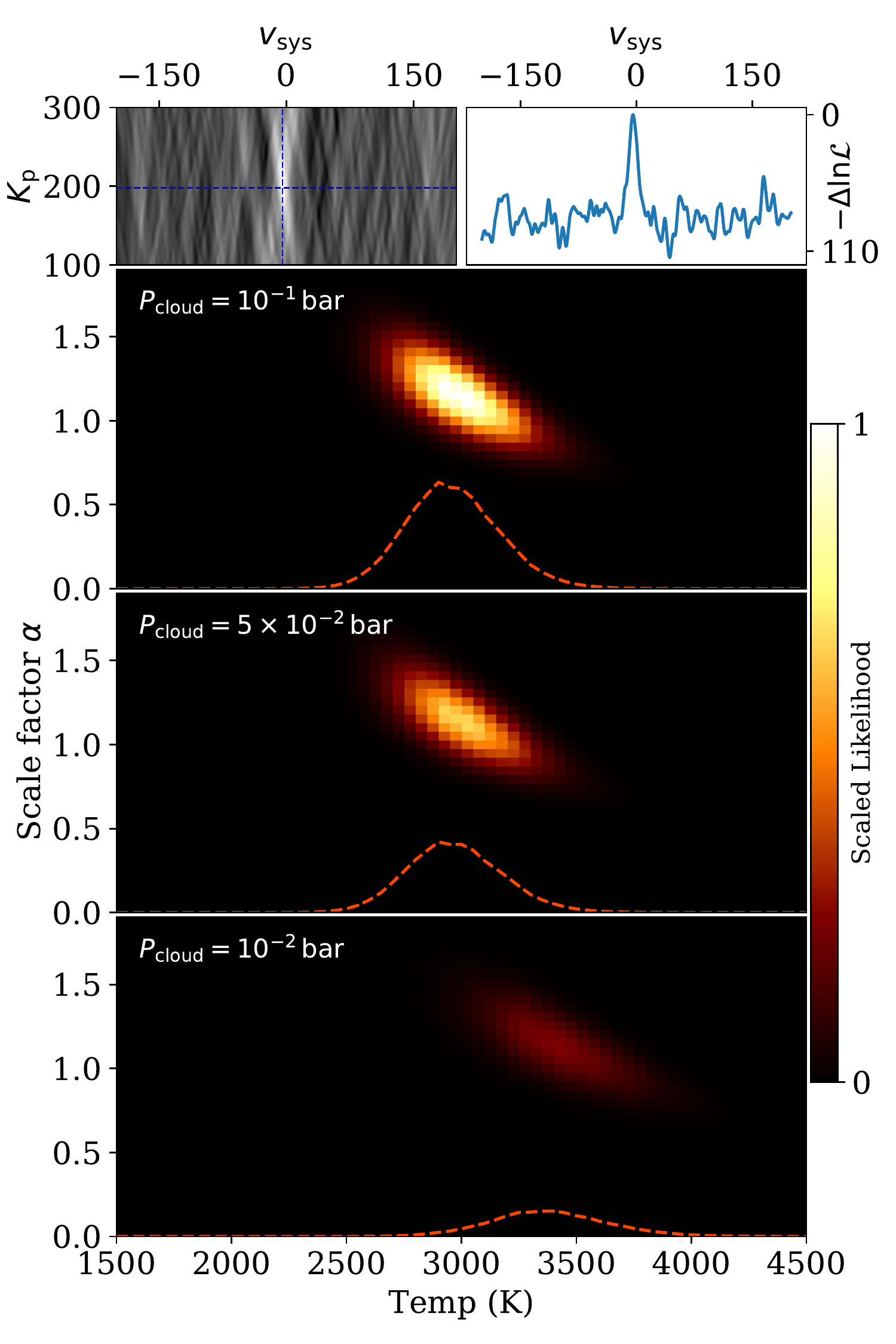}
\caption{Log likelihood maps directly computed from the CCF. Top left: 2D log likelihood map of  $K_\mathrm{p}$ vs $v_\mathrm{sys}$ conditioned on the maximum likelihood of the remaining parameters. Top right: 1D slice of the log Likelihood as a function of $v_\mathrm{sys}$ through the best-fit $K_\mathrm{p}$. Lower panels: Likelihood maps of scale factor $\alpha$ vs model temperature for the maximum values of $K_\mathrm{p}$, and $v_\mathrm{sys}$, for a range of cloud deck pressures as indicated. These maps are all normalised to the maximum likelihood value (in the upper map). The dashed lines shows each map summed (marginalised) over $\alpha$, and normalised to the same scale. Clearly, the likelihood maps constrain the values of $\alpha$ and $T$, but there is a degeneracy with the cloud altitude.
}
\label{fig:logL}
\end{figure}

The results are shown in Fig.~\ref{fig:logL}. The upper panels show the log likelihood map and slice through the best-fit $K_\mathrm{p}$. This is qualitatively indistinguishable from the CCF map. The other three panels show conditional likelihood maps for the best fit $K_\mathrm{p}$ and $v_\mathrm{sys}$ as a function of the scale factor $\alpha$ and model temperature. Each panel shows the results for a different pressure height for the cloud deck. The maps are normalised to the (global) maximum likelihood. Clearly, our data can constrain both the scale and the temperature of the model, enabling direct retrieval of physical parameters from the data. The dotted lines show the (arbitrarily normalised) probability distribution for the temperature after marginalisation over $\alpha$, but conditioned on fixed values of $P_\mathrm{cloud}$, $K_\mathrm{p}$ and $v_\mathrm{sys}$, with typical uncertainties of a few hundred K. Comparing the different maps shows that the values for $\alpha$ and $T$ are correlated with the cloud deck pressure. However, the likelihood prefers the clouds to be located deep in the atmosphere, as the likelihood(s) are larger in the top panel. Thus a full retrieval of temperatures and other parameters from the data must take these degeneracies into account.

\subsection{Fitting the \ion{Fe}{I} feature using a direct likelihood evaluation}
\label{sect:MCMCfits}

Computing full log likelihood maps from the CCFs as a function of $K_\mathrm{p}$ and $v_\mathrm{sys}$, as well as model parameters of the atmospheric model and likelihood quickly becomes prohibitive as the number of parameters increases. Rather than effectively compute the log likelihood for a grid of parameters, it is much more efficient to use Markov Chain Monte Carlo (MCMC) techniques.

To do this, we start with the pre-processed data (after the stellar and telluric signals are removed). The standard technique involves generating a model template, and for each spectrum (and order) interpolating to a grid of velocity shifts $v_\mathrm{sys}$ and integrating over wavelength for {\it every velocity shift} to generate the CCF as a function of $v_\mathrm{sys}$. However, in principle we only need to compute the CCF for a specific $K_\mathrm{p}$ and $v_\mathrm{sys}$ for each likelihood evaluation. We therefore compute the model template, but interpolate it only to the required velocities as a function of time for specific values of  $K_\mathrm{p}$ and $v_\mathrm{sys}$ according to the equation in Sect.~\ref{sect:modelling}. We then simply compute the $\chi^2$ for the model using:
$$
\chi^2 = \sum \frac{(f_i-\alpha m_i)^2}{(\beta\sigma_i)^2}
$$
for given values of $\alpha$ and $\beta$. Finally we evaluate the log likelihood as:
$$
\ln\L = -N\ln\beta-\frac{1}{2} \chi^2,
$$
where we have removed constant terms. The first term involving $\beta$ is now required as we treat $\beta$ as a free parameter.
Our final log likelihood is a function of $K_\mathrm{p}$, $v_\mathrm{sys}$, $\alpha$, $\beta$, $T_\mathrm{atmos}$, $\chi_\mathrm{\ion{Fe}{I}}$, $P_\mathrm{cloud}$, and $\chi_\mathrm{ray}$.
However, we fix $\chi_\mathrm{\ion{Fe}{I}}$ as discussed in Sect.~\ref{sect:modelling}, and initially fix $\chi_\mathrm{ray}=1$ (i.e. the expected H$_2$ Rayleigh scattering assuming Jupiter composition) to directly compare our results to the likelihood maps presented in the previous section. In practice, we fit for the log of $P_\mathrm{cloud}$ and $\chi_\mathrm{ray}$, as this is the natural parameterisation for strictly positive parameters.
It is now straightforward to compute and use this log likelihood directly in optimisation and MCMC routines.

We first determined a global optimum of the posterior using a differential evolution algorithm\footnote{as implemented in the {\sc SciPy} package, based on \citet{Storn_1997}}, and then used a Differential Evolution Markov Chain (DE-MC) algorithm to explore the posterior distribution and extract marginalised distributions \citep{DEMC,Eastman2013} using the implementation described in \citet{Gibson_2019}. We constrained the cloud deck to range between pressures of $10^{-6}$--$1$\,bar, so that the clouds would not completely obscure the \ion{Fe}{I} features, and to restrict degeneracy when the clouds were below the \ion{Fe}{I} or Rayleigh `continuum'. We ran $256$ chains for 400 iterations, discarding the first 40\% of each chain, and confirmed convergence using the Gelman \& Rubin statistic \citep{GelmanRubin_1992} after splitting the chains into four independent groups. We also repeated the procedure to ensure consistency.

\begin{figure*}
\centering
\includegraphics[width=140mm]{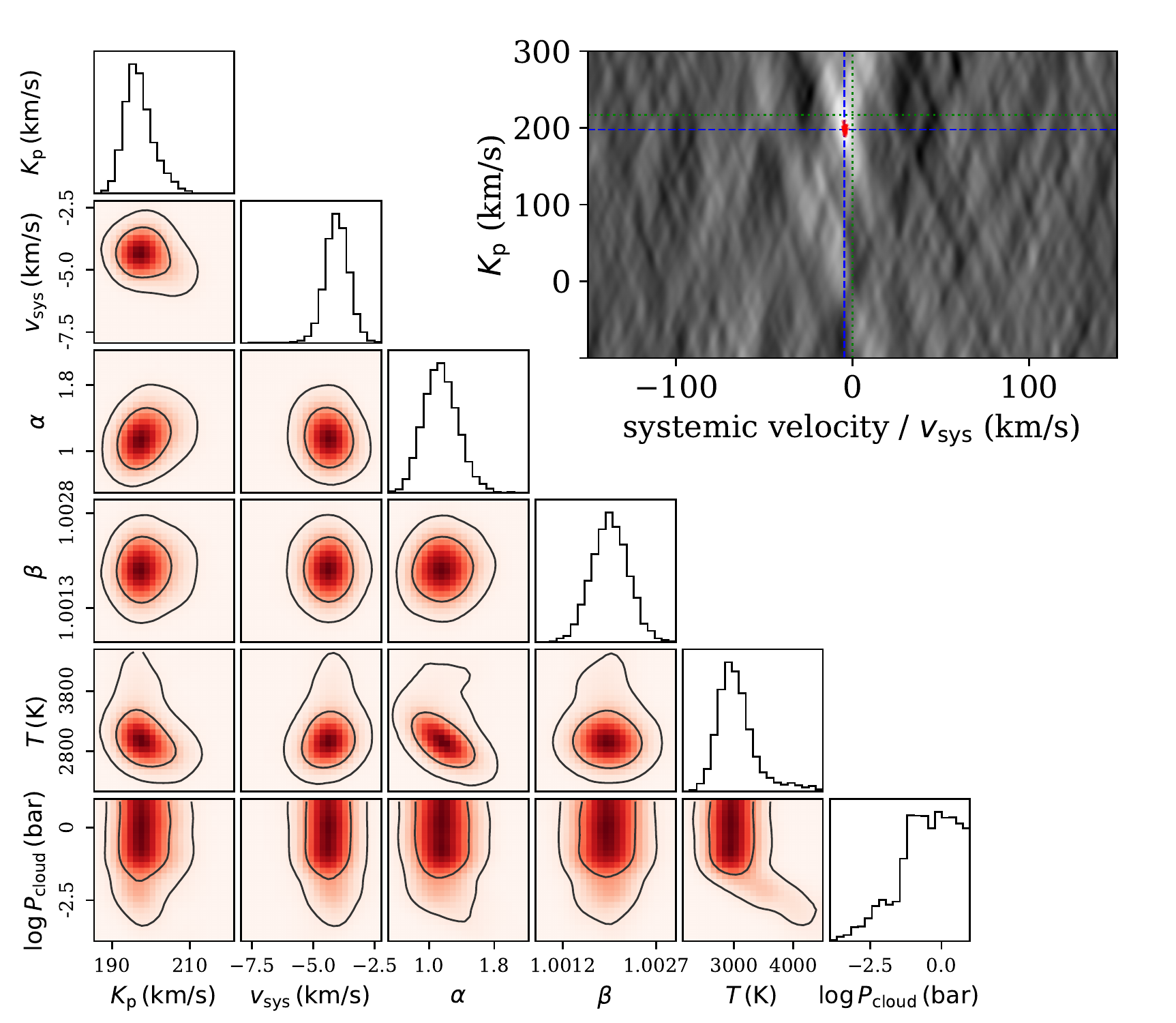}
\caption{Posterior distribution of the parameters of the MCMC fit to the data using an isothermal atmosphere with \ion{Fe}{I} absorption and aerosol absorption as described in Sects.~\ref{sect:modelling} and \ref{sect:MCMCfits}. The \ion{Fe}{I} abundance and Rayleigh scattering parameters are held fixed. Contours mark the 1 and 2\,$\sigma$ limits, respectively. The upper right panel shows the cross-correlation map with samples of $K_\mathrm{p}$ and $v_\mathrm{sys}$ over-plotted, showing the MCMC results in the same peak feature.}
\label{fig:MCMC}
\end{figure*}

The results are shown in Fig.~\ref{fig:MCMC} and the marginalised distributions for each parameter are summarised in Table.~\ref{tab:results}. These results confirm the qualitative picture from the likelihood maps, i.e. that we can constrain the temperature and scale factor of the models, and that there is a strong degeneracy between the cloud deck altitude and the temperature constraints.
{We also repeated the MCMC using the modified \citet{Brogi2019} likelihood function derived in Sect.~\ref{sect:mapping}, which removes dependence on $\beta$ (but still includes time- and wavelength-dependent uncertainties), finding no significant differences. This is expected where we have an extremely high number of data points, and $\beta$ is well constrained by the data. However, as noted in Sect.~\ref{sect:detection_crosscorr}, incorporating the time- and wavelength-dependent uncertainties does make a substantial difference to our results.}

\begin{figure}
\centering
\includegraphics[width=80mm]{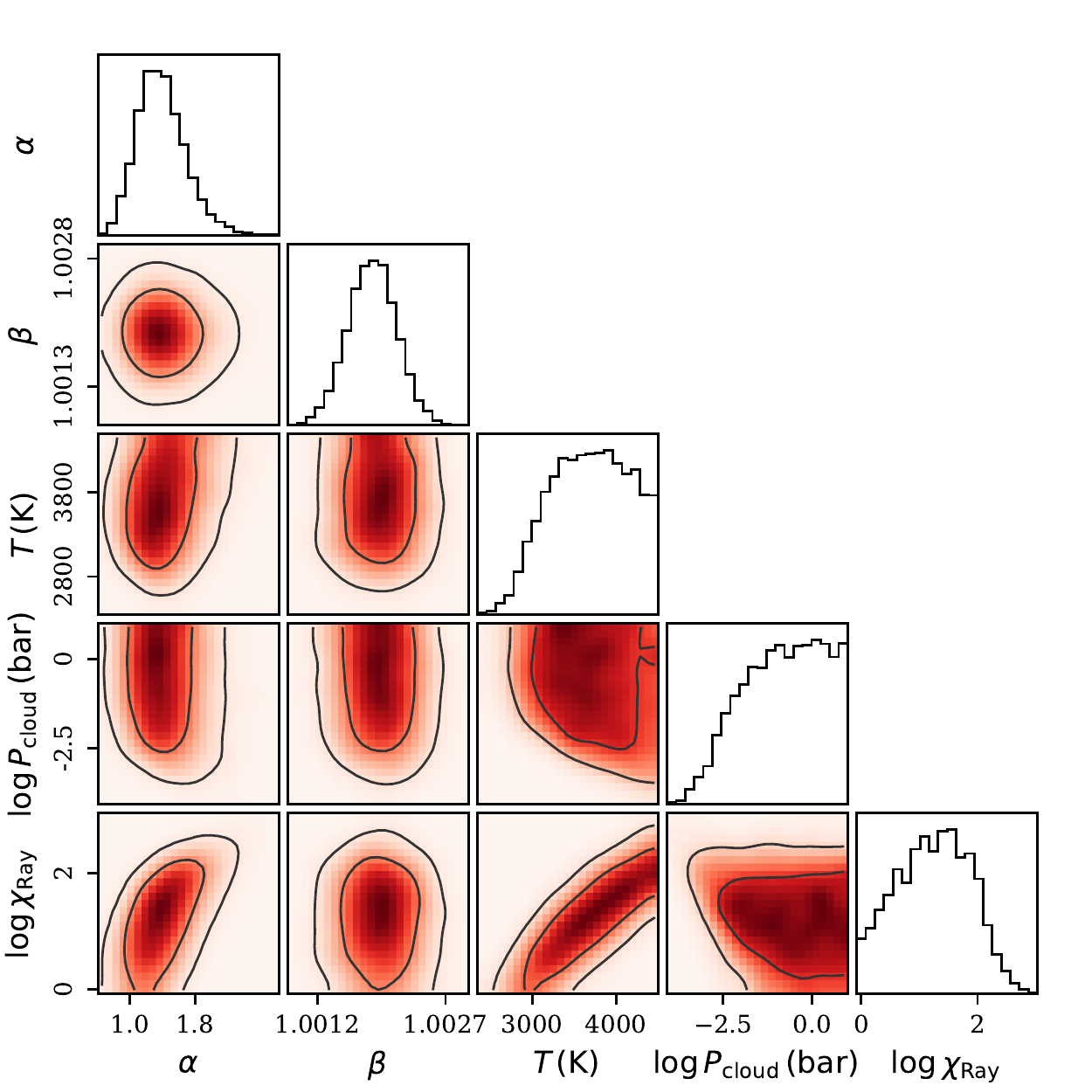}
\caption{Posterior distribution of the parameters of the MCMC fit, similar to Fig.~\ref{fig:MCMC} but with the Rayleigh scattering parameter also free as discussed in Sect.~\ref{sect:MCMCfits}. Note that we also marginalised over $K_\mathrm{p}$ and $v_\mathrm{sys}$, but the resulting distributions are indistinguishable from  Fig.~\ref{fig:MCMC} and are omitted here for clarity.}
\label{fig:MCMC2}
\end{figure}

We finally ran two additional MCMC chains. The first also included the Rayleigh scattering strength as a free parameter, constrained to be between the expected H$_2$ strength (i.e. Jupiter abundance) and an enhancement factor of $10^{4}$, again to avoid totally obscuring any \ion{Fe}{I} lines, similarly to the cloud deck pressure. 
These results are shown in Fig.~\ref{fig:MCMC2} and Table.~\ref{tab:results}.
The second was an MCMC fit with all parameters held fixed at their maximum likelihood values except for $\alpha$ and $\beta$. This was to assess the detection significance of the \ion{Fe}{I} feature, by simply dividing $\alpha$ by its uncertainty, as $\alpha$\,=\,0 would correspond to a flat model (i.e. no detection). We discuss the implications of these results in Sect.~\ref{sect:discussion}. 

\begin{table}
\caption{MCMC results. We report standard deviations were marginalised distributions are symmetrical, and median $\pm$ 1$\sigma$ limits otherwise.}
\label{tab:results}
\begin{tabular}{lll}
\hline
\noalign{\smallskip}
Parameter & Value & Unit\\
\hline
\noalign{\smallskip}
\multicolumn{3}{l}{Fixed Rayleigh scattering:} \\\noalign{\smallskip}
~$K_\mathrm{p}$ & $198.0^{+4.7}_{-3.5}$  & km/s \\\noalign{\smallskip} 
~$v_\mathrm{sys}$ & $-4.39 \pm 0.59$ & km/s \\\noalign{\smallskip} 
~$\alpha$ & $1.16^{+0.22}_{-0.21}$ & - \\\noalign{\smallskip} 
~$\beta$ & $1.00192 \pm 0.00030$ & - \\\noalign{\smallskip} 
~$T$ & $3000^{+310}_{-230}$ & K \\\noalign{\smallskip} 
~$\log P_\mathrm{cloud}$ & $-0.4^{+1.0}_{-1.0}$ & bar \\\noalign{\smallskip} 
\multicolumn{3}{l}{Free Rayleigh scattering:} \\\noalign{\smallskip}
~$K_\mathrm{p}$ & $198.0^{+3.9}_{-3.0}$  & km/s \\\noalign{\smallskip}
~$v_\mathrm{sys}$ & $-4.3\pm{0.56}$ & km/s \\\noalign{\smallskip}
~$\alpha$ & $1.41^{+0.35}_{-0.29}$ & - \\\noalign{\smallskip}
~$\beta$ & $1.00192\pm{+0.00030}$ & - \\\noalign{\smallskip}
~$T$ & $3710^{+490}_{-510}$ & K \\\noalign{\smallskip}
~$\log P_\mathrm{cloud}$ & $-0.8^{+1.3}_{-1.3}$ & bar \\\noalign{\smallskip}
~$\log \chi_\mathrm{ray}$ & $1.27^{+0.64}_{-0.77}$ & - \\\noalign{\smallskip}
\multicolumn{3}{l}{$\alpha$ and $\beta$ only:} \\\noalign{\smallskip}
~$\alpha$ & $1.29\pm0.16$ & - \\\noalign{\smallskip}
~$\beta$ & $1.00192\pm0.00030$ & - \\\noalign{\smallskip}
\hline
\noalign{\smallskip}
\end{tabular}
\end{table}

\subsection{Validation of the new likelihood via injection tests}
\label{sect:injection}

While the framework presented in Sect.~\ref{sect:mapping} is statistically rigorous, the filtering of the data with algorithms such as {\sc SysRem} can in principle alter the underlying planet signal, and potentially lead to biased results. Here, we perform an injection test with a known model in order to validate our method, and verify that the recovered parameters are consistent with the injected signal, in particular the atmospheric temperature ($T_\mathrm{atmos}$) and the scale factor of the model ($\alpha$).

{Rather than perform a complete end-to-end simulation of the data, we chose to inject a signal directly into our data. As the data already contains a real Fe signal, we injected the planetary signal with an identical orbital speed, but with opposite sign. This way, the real and injected signals will overlap as little as possible, but will be treated in much the same way by the data processing algorithms. We injected a signal with $T_\mathrm{atmos}$\,=\,2800\,K, $\alpha$\,=\,3, $P_\textrm{cloud}$\,=\,0.1\,bar, and $\chi_\mathrm{Ray}=1$, with $K_\mathrm{p}=-217$\,km/s and $v_\mathrm{sys}$\,0\,km/s. The forward model was computed as described in Sect.~\ref{sect:modelling}, convolved to the instrumental resolution, and converted to fractional absorption as a function of time after being weighted by the transit model of the planet (assuming no limb darkening). We finally Doppler-shifted the spectra according to the planet's velocity and interpolated to the wavelengths of each spectral order before injecting into the data. The signal was injected after the blaze correction, and was processed through {\sc SysRem} with 15 iterations as before. We expect that the earlier steps (cleaning and blaze correction) would have negligible impact on the injected signal.
}

\begin{figure*}
\centering
\includegraphics[width=155mm]{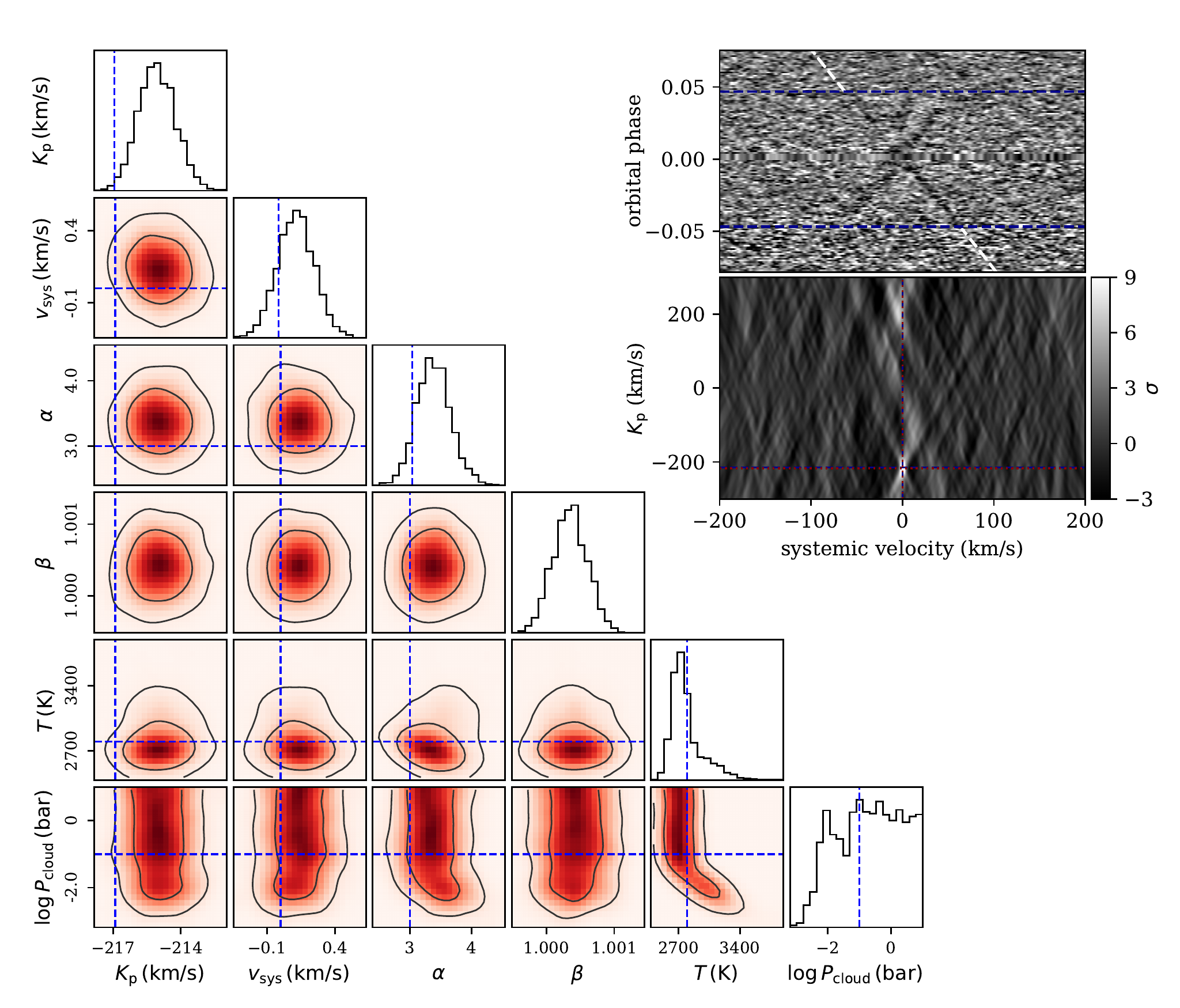}
\caption{Results from the injection test. Left: Posterior distribution of the parameters in the MCMC fit, with the value of the Rayleigh scattering parameter held fixed. The dashed lines mark the known values of the injected model (note that there is no known value for $\beta$). Upper right: Summed cross-correlation of the bluest 13 orders, and velocity-summed cross-correlation map, analogous to the plots shown in Fig.~\ref{fig:processing}. Note that both the real and injected signals are visible in both plots.}
\label{fig:injection}
\end{figure*}

{
We then proceed to run the same algorithms as before, including the CCF and likelihood maps, and MCMC analysis. In this case we tested additional methods to filter the template model, including a high-pass filter (implemented as a Butterworth filter using {\sc scipy.signal.filtfilt}), as well as various filter widths/frequencies. The results are summarised in Fig.~\ref{fig:injection} (here using a high-pass filter), demonstrating that the recovered signal is consistent with the injected model. For all parameters we found that the recovered values agreed with the injected signal within 1\,$\sigma$, with the exception of $K_\mathrm{p}$ which is within 2\,$\sigma$. We found that the various model-filtering methods (or chosen filter widths, filtering frequencies, within reason) did not dramatically influence our results more than 1\,$\sigma$.}

{We can therefore conclude that the extracted values from our WASP-121b data are not significantly impacted by our model-filtering methods. Nonetheless, it is clear that the data processing and filtering of the model can in principle influence the recovered signal, and this injection test only shows that our chosen techniques work at the approximate level of precision reported here for our Fe I signal in WASP-121b. We do not claim that the filtering methods we apply are fully optimised, or are the best approach for high-resolution observations. A complete exploration of this would require more sophisticated atmospheric models to explore a wider range of atmospheric parameters, as well as different levels of signal-to-noise and various filtering-algorithms to process the models. Understanding the impact of the removal of telluric and stellar signals on the planet's atmospheric signal, and therefore how to filter the atmospheric models accordingly, is arguably the most important remaining challenge in fully exploiting the high-resolution technique for probing exoplanet atmospheres. We refer the reader to \citet{Cabot2019} for a discussion of the removal of stellar and telluric signals using various algorithms, or \citet{Brogi2019} in the context of Bayesian retrieval. {The latter perform a full end-to-end simulation of a high-resolution time-series dataset, which allows a more detailed exploration of the impact of different levels of noise and noise structures on our methods, whereas our approach only explores a specific realisation of the noise (i.e. from our observed dataset).} A full exploration of the model-filtering and data pre-processing may be the subject of future work. Finally, we note that our injection test does not account for the effects of the RM effect. While this is unlikely to influence the recovery of WASP-121b's atmospheric signal, it is of importance for fast-rotating host stars.
}

\section{Discussion}
\label{sect:discussion}

Our results demonstrate a clear detection of \ion{Fe}{I} in the atmosphere of WASP-121b, using both the standard cross-correlation technique and a new likelihood-based approach. Using the cross-correlation approach and a simple estimate of the noise, we measure a detection significance of 9.2\,$\sigma$. Using our MCMC fits to the data directly, the detection significance becomes 8.2\,$\sigma$. We also detect the feature near the expected $v_\mathrm{sys}$ and $K_\mathrm{p}$, giving further confidence in our results, in addition to the fact that no other significant peaks appear in the CCF map.
Fe detections have previously been reported for a range of ultra-hot Jupiters \citep[e.g.][]{Haswell2012,Hoeijmakers2018,Hoeijmakers2019,Casasayas-Barris2019}, with \ion{Fe}{i} in particular recently predicted to be an important atmospheric constituent in hot atmospheres \citep[e.g.][]{Kitzmann2018,Lothringer2018,Lothringer2019}.

\ion{Fe}{II} was also recently found in the exosphere of WASP-121b through UV transmission spectroscopy \citep{Sing2019}.
Previous observations of WASP-121b at optical and near-infrared wavelengths detected a temperature inversion through direct measurement of water emission features \citep{Evans2017}. This was naturally considered to be further evidence for the presence of TiO or VO in the atmosphere \citep[for a long time the leading suspected cause of thermal inversions in hot-Jupiters;][]{Fortney_2008}, but subsequent observations have been unable to confirm this \citep{Evans2018,Evans2019}. Our observations show that \ion{Fe}{I} is the most likely contributor to the temperature inversion detected in the stratosphere. {Given that the planet exhibits strong absorption at blue-optical wavelengths, and that it orbits an F6-star}, a significant amount of energy would easily be deposited at high altitudes. Previously detected ionised species such as \ion{Fe}{II} and \ion{Mg}{II} likely deposit energy much higher in the atmosphere. However, this does not rule out the presence of additional optical absorbers that may also contribute; in particular we are yet to verify the cause of absorption features seen at $\approx$600--800\,nm in low-resolution transmission spectra \citep{Evans2018}.

Our observations also detect a slight offset of the \ion{Fe}{I} feature in $v_\mathrm{sys}$. This is apparent in both the CCF and likelihood maps, as well as the MCMC fits. The offset of $-4.4\pm0.6\,\mathrm{km/s}$ is formally a significant detection, and we confirmed the stellar signal is centred at $v_\mathrm{sys}$\,=\,0 by following the same cross-correlation process with the \ion{Fe}{I} model without removing the stellar signals (rather we continuum subtracted after the blaze correction). The ephemeris of the planet is also precise enough to rule out phase offsets in the transit centre causing the offset (an offset of $\approx$7\,mins is required to cause the shift in $v_\mathrm{sys}$). 
Physically, this is plausibly the result of winds in the planet's atmosphere, predicted to be of order $\approx$5\,{km/s} for ultra-hot Jupiters \citep[e.g.][]{Kataria2016}. A detection of this magnitude would require either significant asymmetry in the absorption (e.g. significant high-altitude cloud coverage obscuring the red-shifted absorption from the atmosphere, spatially varying abundances), or a bulk flow of material from the dayside to the nightside at the altitudes probed by our observations. However, while this is an intriguing signal, doubts remain about the wavelength stability of UVES. In particular the wavelength correction for each order were inconsistent at the km/s level, demonstrating difficulties with the wavelength solution. While we corrected for wavelength shifts per order, we did not refine the determination of the dispersion in each order, and is difficult to perform reliably with image-slicer observations where the trace position is harder to locate accurately. Therefore we would caution against physical interpretation of this offset (or of the value of $K_\mathrm{p}$), and further observations are needed to confirm or refute this result.

We have also developed a new likelihood-based method for extracting physical parameters from high-resolution data, largely motivated by the work of \citet{Brogi2019}. Our method has a number of advantages, including being able to determine the scaling of the model features, {explicitly} account for the time- and wavelength-dependent noise (with a noise scale factor), as well as constraining atmospheric model parameters. The fits for the scale factor parameter $\alpha$ show that the underlying model is best fit using a scale factor only marginally greater than the model atmospheres generated assuming a scale height of 960\,km (i.e. $\alpha\gtrsim1$). As we fixed the temperature to 2,800\,K for the scale height calculation, it is possible that a slightly higher temperature could explain this larger amplitude signal. Alternatively, the dissociation of H$_2$ could lead to a lower mean molecular weight \citep[e.g.][]{Parmentier2018,Lothringer2018}, similarly increasing the scale height assumed in our calculation.

However, these are relatively small effects. Our results immediately show that the \ion{Fe}{I} signal is largely constrained to relatively deep within the atmosphere, i.e. at the levels probed by previous low-resolution optical and near-infrared transmission spectra.
In contrast, the \ion{Fe}{II} feature detected by \citet{Sing2019} extends up to more than twice the planetary radius. This would correspond to a scale factor of $\gg$10, clearly showing that the \ion{Fe}{I} feature originates in a physically distinct region from \ion{Fe}{II}.
In our model we find that \ion{Fe}{I} is located deeper than $\sim$1$\,\upmu$bar when fixing the volume mixing ratio to 10$^{-6}$, but this is degenerate with the abundances and scattering properties. This is consistent with the high temperatures in the upper atmosphere measured by \citet{Sing2019}, and the prediction that \ion{Fe}{I} will be strongly ionised in the upper atmosphere \citep[e.g.][]{Lothringer2018,Lothringer2019}. These results also suggest that Fe is almost completely (singly) ionised in the upper atmosphere, otherwise we would expect to see a much larger \ion{Fe}{I} signal. We also searched for \ion{Fe}{II} features in our data, but were unable to detect a significant signal. This could be explained by the fact that there are significantly fewer strong \ion{Fe}{II} lines than \ion{Fe}{I} at blue optical wavelengths (particular at the high temperatures in the exosphere), which could lead to a weaker signal. We could expect the extended exosphere to compensate for this, as lines in the exosphere would extend more than $\gg$10 times depth measured here as shown by the \ion{Fe}{II} lines. Another possibility is that the \ion{Fe}{II} lines are significantly broadened due to a larger velocity range within the extended (and escaping) exosphere, which would be removed during the pre-processing of high-resolution data. It is also possible that an escaping atmosphere does not neatly follow the same orbit as the planet, which could complicate detection at high-resolution, yet not be obvious from the UV observations. This potential for smearing out the lines high in the exosphere could in principle also hide some remnant neutral Fe present in the upper atmosphere, which makes it difficult to completely rule out its presence in the exosphere, although \ion{Fe}{i} lines were not detected in the UV by \citet{Sing2019}.

In addition, we constrained the temperature of the atmosphere. We determine a temperature of $3000^{+310}_{-230}$\,K with the cloud deck level left as a free parameter and fixing the Rayleigh scattering feature, or $3710^{+490}_{-510}$\,K when allowing both scattering parameters to be free. Clearly, the temperature is highly degenerate with the scattering properties. This is unsurprising, as the temperature is determined from the relative strength of lines rising above the continuum. Furthermore, assuming an isothermal atmosphere is a poor assumption in the case of WASP-121b; however, extending our MCMC approach to account for a varying temperature-pressure profile will require more efficient forward models of the atmosphere. We also do not consider the effect of pressure broadening. This is perhaps a reasonable assumption, given that we are typically extracting information high in the atmosphere with high-resolution transmission spectroscopy, although exploring pressure sensitivity may potentially provide valuable information to break degeneracies in abundances and scattering properties. We re-emphasise that the abundances, pressure level of the clouds, and relative Rayleigh scattering strength are degenerate in our model, and should not be interpreted in a physically meaningful way. We aim to explore these issues in future work.

Nonetheless, our temperature constraints suggest that the (averaged) temperature in the regions probed by our observations is higher than that detected in previous studies of both the transmission and emission spectrum. Given that high resolution observations probe higher layers in the atmosphere, and that WASP-121b has been found to exhibit a temperature inversion, a higher temperature is consistent with expectations. Finally, the UV observations and detection of \ion{Fe}{II} by \citet{Sing2019} imply a temperature of $>$5,000\,K, again consistent with \ion{Fe}{I} being present deeper in the atmosphere (at cooler temperatures) and \ion{Fe}{II} only in the extended, hotter exosphere.

\section{Conclusions}
\label{sect:conclusion}

We have presented a clear detection of \ion{Fe}{I} in the atmosphere of the ultra-hot Jupiter WASP-121b at $>$$8$\,$\sigma$. WASP-121b was previously shown to host a temperature inversion, and \ion{Fe}{I} is the first unambiguously detected species that could be a major absorber of incoming optical radiation from the host star that could deposit enough energy in the stratosphere. We use a new likelihood-based approach that is able to fit for a scaling factor of the model atmosphere, as well as recover the temperature of the atmosphere after marginalising over scattering properties. We recover a temperature substantially hotter than that previously detected using low-resolution observations, consistent with the temperature inversion and our expectation that high-resolution observations probe higher altitudes. 
We also show that the \ion{Fe}{I} feature originates from relatively deep within the atmosphere, i.e. that the origin of the \ion{Fe}{I} feature is physically distinct form the extended \ion{Fe}{II} feature previously detected at UV wavelengths, consistent with the expectation that any Fe in the exosphere is strongly ionised. The fact that we do not detect \ion{Fe}{I} in the exosphere suggests that Fe in the exosphere is fully (singly) ionised.

Finally, we have demonstrated that high-resolution time-series observations can be interpreted via direct likelihood evaluations, using similar statistical techniques to low-resolution transit observations, and complementary to several new approaches to high-resolution analysis \citep[e.g.][]{Pino2018b,Brogi2019,Watson2019,Fisher2019}. Cross-correlation will likely remain the most efficient technique for detecting atomic and molecular species using high-resolution observations, but these new methods are capable of comparing model fits and therefore extracting posterior probability distributions of physically meaningful parameters of the planet's atmosphere. Specifically, our method (to our knowledge) is the first to demonstrate the use of a standard Gaussian likelihood function on high-resolution Doppler-resolved observations (i.e. without modification), and also enables the extraction of a scale factor (to extract the extent of the model atmosphere, of particular use for modelling extended signals from ultra-hot Jupiters), as well as allowing the inclusion of time- and wavelength-dependent uncertainties. We demonstrated our technique using simple, isothermal atmospheres, and ignoring pressure broadening. The use of more sophisticated models should enable more stringent constraints on the planet's atmosphere, and we expect that with the detection of multiple species we could place strong limits on relative abundances. Furthermore, this method can trivially be coupled with low-resolution observations, either by combining posterior distributions, or via joint fits \citep[similarly to that demonstrated in][]{Brogi2019}. Future work will explore further innovations using a direct likelihood evaluation for high-resolution Doppler-resolved spectroscopy.

\section*{Acknowledgements}

We are extremely grateful to the referee, Matteo Brogi, for careful reading of the manuscript. This work is based on observations collected at the European Organisation for Astronomical Research in the Southern Hemisphere under ESO programme 098.C-0547. N. P. G. gratefully acknowledges support from Science Foundation Ireland and the Royal Society in the form of a University Research Fellowship. S.K.N and C.A.W. would like to acknowledge support from UK Science Technology and Facility Council grant ST/P000312/1.
This work also benefited from discussions at the 2019 Exoplanet Summer Program held at the Other Worlds Laboratory (OWL) at the University of California, Santa Cruz, a program funded by the Heising-Simons Foundation. 
We are grateful to the developers of the {\sc NumPy, SciPy, Matplotlib, iPython} and {\sc Astropy} packages, which were used extensively in this work \citep{Jones_2001,Hunter_2007,Perez_2007,Astropy}.

%




\bibliography{/Users/ng//MyDocuments/MyBibliography} 
\bibliographystyle{mnras} 



%
%


\bsp	
\label{lastpage}
\end{document}